\documentclass[prl,twocolumn,aps,longbibliography,superscriptaddress,notoc]{revtex4-2}
\usepackage{bm}
\usepackage{graphicx}
\usepackage{amssymb}
\usepackage{amsmath}
\usepackage{eufrak}
\usepackage{color}
\usepackage[utf8]{inputenc}
\usepackage{ulem}
\usepackage[unicode=true,colorlinks=true,citecolor=blue,urlcolor=blue]{hyperref}

\renewcommand{\Re}{\mathop{\rm Re}}
\renewcommand{\Im}{\mathop{\rm Im}}

\newcommand{\ch}{\mathop{\rm ch}}
\newcommand{\sh}{\mathop{\rm sh}}

\newcommand{\arctg}{\mathop{\mathrm{arctg}}}
\let\ifr\i
\renewcommand{\i}{{\rm i}}
\renewcommand{\d}{\mathrm d}
\renewcommand{\emph}{\textit}
\newcommand{\braket}[1]{\left\langle #1 \right\rangle}

\usepackage{chngcntr}
\newcommand{\enquote}{}
\graphicspath{{img/}}

\newcommand{\nix}[1]{}
\let\oldsec\section
\renewcommand{\section}[1]{\textit{#1}---}

\begin{document}

\title{Kondo enhancement of current induced spin accumulation in a quantum dot}

\author{V.~N.~Mantsevich}
\affiliation{Chair of Semiconductors and Cryoelectronics and Quantum Technology Center, Faculty of Physics, Lomonosov Moscow State University, 119991 Moscow, Russia}
\author{D.~S.~Smirnov}
\email[Electronic address: ]{smirnov@mail.ioffe.ru}
\affiliation{Ioffe Institute, 194021 St. Petersburg, Russia}

\begin{abstract}
  Weak spin-orbit coupling produces very limited current induced spin accumulation in semiconductor nanostructures. We demonstrate a possibility to increase parametrically the spin polarization using the Kondo effect. As a model object we consider a quantum dot side coupled to a quantum wire taking into account the spin dependent electron tunneling from the wire to the dot. Using the nonequilibrium Green's functions, we show that the many body correlations between the quantum dot and the quantum wire can increase the current induced spin accumulation at low temperatures by almost two orders of magnitude for the moderate system parameters. The enhancement is related to the Kondo peak formation in the density of states and the spin instability due to the strong Coulomb interaction. This effect may be useful to electrically manipulate the localized electron spins in quantum dots for their quantum applications.
\end{abstract}

\maketitle{}

\section{Introduction}Semiconductor quantum dots (QDs) hold a great promise for the scalable quantum information processing using the localized spins in QDs as qubits~\cite{Michler2017,Kloeffel2013,Watson2018,Yoneda2018,Yang2020}. The electron and hole spins can be efficiently oriented~\cite{PhysRevLett.94.047402,PhysRevLett.99.097401,gerardot08}, manipulated~\cite{greilich06,press08,Greilich2009,Dusanowski2022} and read out~\cite{berezovsky2006nondestructive,atature07,Arnold2015} by optical means. However, electrical schemes being based on the highly advanced fabrication technology suggest larger ensembles of individually addressable qubits. On this way, the electrical spin transport, spin correlations, and spin read out are already firmly established~\cite{Nowack07,ShchepetilnikovNMR,Zajac439,Mills2019,PhysRevLett.126.017701,PhysRevLett.126.107401}. Only the electrical single spin polarization without magnetic field remains elusive for years.

This stumbling block on the way of quantum technologies can be removed by the current induced spin polarization effect~\cite{dyakonov_book}. The nonequilibrium flow of charge carriers breaks the time inversion symmetry and allows for the polarization of electron spins in the system. This effect was first predicted~\cite{ivchenko1978new} and observed~\cite{vorob1979optical} in bulk Te crystals. Later it was extended to the epilayers~\cite{PhysRevLett.93.176601} and quantum wells~\cite{Ganichev_110,silov04} based on GaAs-like semiconductors. Nowadays the related effect of chirality induced spin selectivity is in the focus of intense theoretical and experimental investigations~\cite{Yang2021,Kim2021,Evers2022}.

Generally, the degree of current induced spin polarization is small. This is related to the weakness of the spin-orbit coupling and the small ratio of the drift and Fermi velocities~\cite{ganichev2012spin}. A number of approaches to overcome these factors were suggested such as: spin-momentum locking~\cite{Li2016,Vaklinova2016,Tian2017}, streaming conductivity regime~\cite{Golub2013,Golub2014}, hopping conductivity~\cite{Hopping_spin,Sherman_QW}, and exploitation of the valence band spin-orbit splitting~\cite{murakami03,Mantsevich2022}. In this Letter, we demonstrate that the spin polarization can be drastically increased at low temperatures due to the Kondo many body correlations. Notably, this Kondo enhancement of the spin accumulation can be combined with the previously established tools to obtain the largest spin polarization.

\begin{figure}
  \centering
  \includegraphics[width=0.95\linewidth]{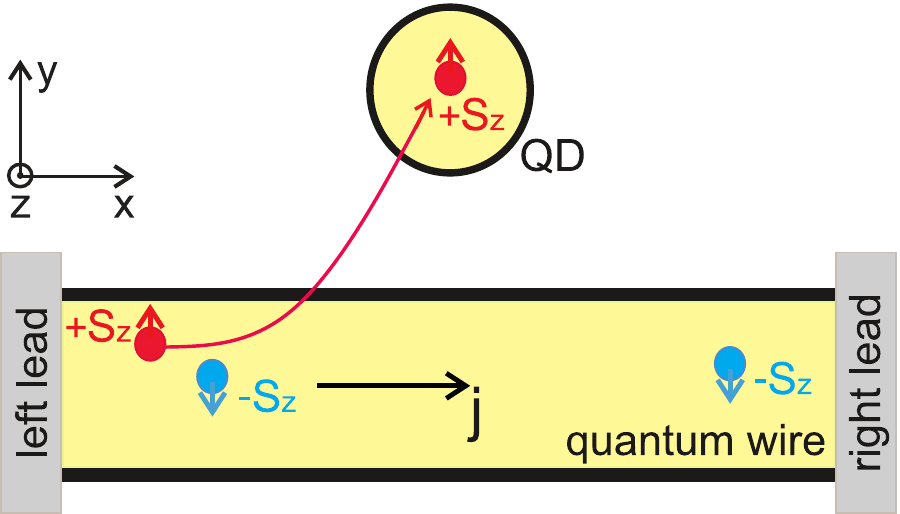}
  \caption{Sketch of a QD side coupled to a quantum wire. The difference of the tunneling probabilities for spin up (red balls with arrows) and spin down (blue ones) particles leads to the current induced spin accumulation in the QD.
  }
  \label{fig:system}
\end{figure}

As a model system, we consider a QD side coupled to a quantum wire, see Fig.~\ref{fig:system}. The structure is assumed to be gate defined in two dimensional electron or hole gas. The spin-orbit interaction gives rise to the spin dependent tunneling. It leads to the spin accumulation in the QD under current flow in the quantum wire~\cite{Mantsevich2022}, similar to the spin Hall~\cite{Sinova2015} and Mott~\cite{abakumov72} effects. At the same time, the Coulomb interaction in the QD produces the many body correlations and leads to the Kondo effect~\cite{hewson1997kondo}. This effect is known to enhance conductivity and the spin susceptibility to external magnetic field~\cite{Krishnamurthy1980}. Here, we demonstrate also the enhancement of the current induced spin accumulation effect.


\section{Model}The $C_{2v}$ point symmetry group of the system allows for the linear coupling between the $x$ component of a vector and the $z$ component of a pseudovector, i.e. between the electric current along the wire and the spin polarization in the QD along the growth axis, see Fig.~\ref{fig:system}. To describe this coupling microscopically, we adopt the Anderson Hamiltonian~\cite{PhysRev.124.41}:
\begin{multline}
  \label{eq:Ham_main}
  \mathcal H=E_0\sum_\pm n_\pm+Un_+n_-+\sum_{k,\pm}E_kn_{k,\pm}\\
  +\sum_{k,\pm}\left(V_{k,\pm}d_\pm^\dag c_{k,\pm}+{\rm H.c.}\right).
\end{multline}
Here $E_0$ is a single particle energy in the QD, $n_\pm=d_\pm^\dag d_\pm$ are the occupancies of the corresponding spin up and spin down states expressed through the annihilation operators $d_\pm$, $U$ is the Coulomb repulsion energy, $E_k$ describes the dispersion of particles in the quantum wire with the wave vector $k$ along the wire, and $n_{k,\pm}=c_{k,\pm}^\dag c_{k,\pm}$ are the occupancies of the corresponding spin states expressed through the annihilation operators $c_{k,\pm}$. We assume the wire to be ballistic and also neglect the interactions in it. Finally, the coefficients $V_{k,\pm}$ describe the spin dependent tunneling between the quantum wire and the QD. Note, that the spin dependence in the form $V_{k,+}\neq V_{k,-}$ is allowed for any crystal structure of the host semiconductors, so the current induced spin accumulation is equally possible for GaAs, Si, and Ge-based heterostructures. The time reversal symmetry imposes a relation $V_{k,+}=V_{-k,-}^*$.

The spin-orbit coupling can lead to the spin splitting of the electron dispersion in the quantum wire. This however requires low symmetry of the system compared to the spin dependent tunneling and is not important for the current induced spin accumulation in the QD separated from the quantum wire by a tunnel barrier. The ratio of $V_{k,+}$ and $V_{k,-}$ determines the chirality of the quasi bound state in the QD~\cite{lodahl2017chiral,PhysRevB.98.235416,Spitzer2018,PhysRevLett.126.073001}. For completely chiral quasi bound state $V_{k_0,-}=V_{-k_0,+}=0$ and $\left|V_{k_0,+}\right|=\left|V_{-k_0,-}\right|\neq0$, so the spin orientation in the QD is locked to the propagation direction along the quantum wire (here $k_0>0$ is determined by the relation $E_{k_0}=E_0$). We have shown recently, that this limit can be reached in the heterostructures with the two dimensional hole gas due to the complex valence band structure~\cite{Mantsevich2022}. In what follows, we focus mainly on the completely chiral states and discuss the case of finite chirality in the end of the Letter.


\section{Formalism}For the calculation of the current induced spin accumulation in the QD as a function of the bias applied to the quantum wire, we use the nonequilibrium Green's functions~\cite{RevModPhys.58.323,stefanucci_vanleeuwen_2013,Arseev_2015}. This approach despite having disadvantages~\cite{Kashcheyevs2006}, allows us to account for the Kondo effect even beyond the linear response regime in the simplest way and to demonstrate the Kondo enhancement of current induced spin accumulation. The truncation of the system of the Heisenberg equations of motion allows one to obtain a closed set of equations for the operators $d_\pm$, $c_{k,\pm}$, $d_\pm n_\mp$, $c_{k,\pm}n_\mp$, $c_{k,\mp}^\dag d_\mp d_\pm$, and $c_{k,\mp}d_\mp^\dag d_\pm$. From its solution one finds the retarded Green's functions of spin up and spin down particles in the QD $G_\pm^R(\omega)$. In particular, in the limit of the large Coulomb repulsion, $U\to\infty$, one obtains ($\hbar=1$)~\cite{Lacroix_1981,PhysRevLett.70.2601,PhysRevB.68.195318}
\begin{equation}
  G_\pm^R(\omega)=\frac{1-\braket{n_\mp}}{\omega-E_0-\Sigma_{0}(\omega)-\Sigma_{1,\pm}(\omega)},
  \label{eq:GR_main}
\end{equation}
where $\braket{n_\mp}$ are the average occupancies of the QD states, which should be determined self consistently. The self energies $\Sigma_{0}(\omega)$ and $\Sigma_{1,\pm}(\omega)$ in the wide band approximation have the form
\begin{subequations}
  \begin{equation}
    \Sigma_{0}(\omega)=\frac{\Gamma}{\pi}\int\limits_{-W}^W\frac{1}{\omega-E+\i 0}\d E,
  \end{equation}
  \begin{equation}
    \Sigma_{1,\pm}(\omega)=\frac{\Gamma}{\pi}\int\limits_{-W}^W\frac{f_{R/L}(E)}{\omega-E+\i 0}\d E.
  \end{equation}
\end{subequations}
Here $\Gamma$ is the tunneling rate between the QD and the quantum wire, $W$ is the band width, and
\begin{equation}
  \label{eq:fLR}
  f_{L/R}(E)=\frac{1}{1+\exp\left[(E-E_F^{L/R})/T\right]}
\end{equation}
are the Fermi distribution functions in the left and right leads with $T$ being the temperature ($k_B=1$), and $E_F^{L/R}$ being the Fermi energies in the left/right leads. Eq.~\eqref{eq:GR_main} is qualitatively correct in the weak coupling regime, when the temperature exceeds the Kondo temperature $T_K=W\exp\left(-\pi|E_F-E_0|/\Gamma\right)$.

The density of states related to the QD is given by $D(\omega)=-\Im\left[G_+^R(\omega)+G_-^R(\omega)\right]/\pi$. It is shown in Fig.~\ref{fig:Polarization_eV_DOS}(a) for the thermal equilibrium, $E_F^L=E_F^R$, and different temperatures. Generally it consists of a broad peak with the width $\sim\Gamma$ at the QD energy $E_0$ (which is a bit renormalized due to the tunneling) and a narrow peak at the Fermi energy, which leads to the Kondo effect. The peak has a width of the order of $T_K$ and disappears with increase of the temperature, as one can see in Fig.~\ref{fig:Polarization_eV_DOS}(a).

To calculate the occupancies of the spin states in the QD, we consider the expressions for the current to the QD from the left/right lead corresponding to spin up/down particles~\cite{haug2008quantum}:
\begin{equation}
  J_{L/R}=e\Gamma\int\frac{\d\omega}{2\pi}\left[\i G_{+/-}^<(\omega)-2f_{L/R}(\omega)\Im G_{+/-}^R(\omega)\right],
\end{equation}
where $G_\pm^<(\omega)$ are the lesser Green's functions of the QD. In the steady state, these currents vanish, which yields the desired occupancies:
\begin{multline}
  \braket{n_{+/-}}=-\i\int\frac{\d\omega}{2\pi}G_{+/-}^<(\omega)\\=-\int\limits_{-W}^W\frac{\d\omega}{\pi}f_{L/R}(\omega)\Im G_{+/-}^R(\omega).
\end{multline}
Here the retarded Green's functions also depend on the occupancies, Eq.~\eqref{eq:GR_main}, so they should be calculated self consistently. Ultimately, the spin polarization in the QD is given by
\begin{equation}
  \label{eq:P_main}
  P=\frac{\braket{n_+}-\braket{n_-}}{\braket{n_+}+\braket{n_-}}.
\end{equation}
Its calculation and analysis is the main goal of this work.

\section{Results}
\begin{figure}
  \centering
  \includegraphics[width=0.95\linewidth]{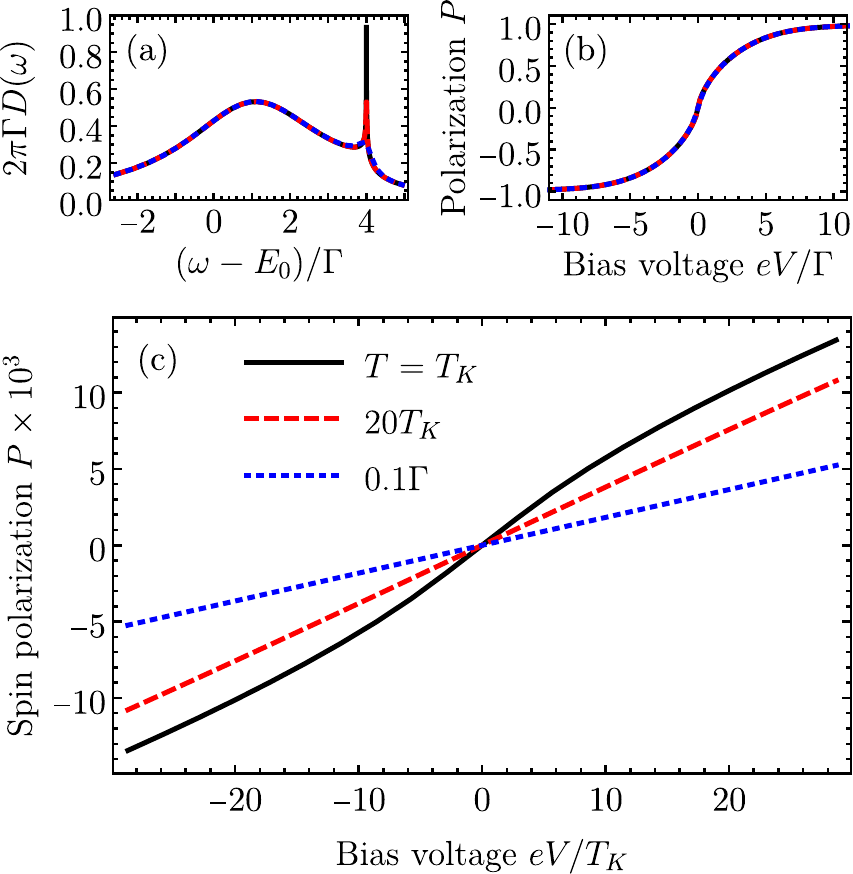}
  \caption{(a) Density of states related to the QD calculated after Eq.~\eqref{eq:GR_main}. (b) Degree of spin polarization in the QD [Eq.~\eqref{eq:P_main}] as a function of the applied bias. (c) The same as in (b) for smaller voltages. The solid black, red dashed and blue dotted curves for all panels are calculated for the different temperatures given in the legend in (c). The other parameters are $E_F-E_0=4\Gamma$ and $W=100\Gamma$.}
  \label{fig:Polarization_eV_DOS}
\end{figure}

The spin polarization induced by the electric current is shown in Fig.~\ref{fig:Polarization_eV_DOS}(b) as a function of the bias $eV$, which is applied symmetrically: $E_F^{L/R}=E_F\pm eV/2$. This dependence looks the same for the three temperatures $T=T_K$, $20T_K$ and $0.1\Gamma$ shown in the figure. Qualitatively, the bias produces the difference of the fluxes of the particles moving to the left and to the right along the quantum wire, and since the tunneling matrix elements depend on the spin and direction of the propagation, this results in the current induced spin accumulation in the QD. At the same time, the large bias $eV\sim\Gamma\gg T_K$ destroys many body correlations, so the dependences shown in Fig.~\ref{fig:Polarization_eV_DOS}(b) overlap for the different temperatures. The spin polarization is an odd function of the bias in agreement with the time reversal symmetry. It increases with increase of the bias and saturates at $|eV|>2|E_F-E_0|$. The polarization approaches 100\% when the quasi bound state in the QD with the width $\Gamma$ lies completely below the Fermi energy in one lead and above the Fermi energy in the other lead.


The many body correlations, which lead to the Kondo effect, manifest themselves at the smallest voltages, $eV\sim T_K$, as shown in Fig.~\ref{fig:Polarization_eV_DOS}(c). Here the same curves as in Fig.~\ref{fig:Polarization_eV_DOS}(b) are zoomed in. One can see, that the current induced spin accumulation is in fact temperature dependent for the smallest voltages: the higher the temperature, the smaller the polarization. For the large voltages the curves for the low temperatures approach the curve for the high temperature.

\begin{figure}
  \centering
  \includegraphics[width=0.95\linewidth]{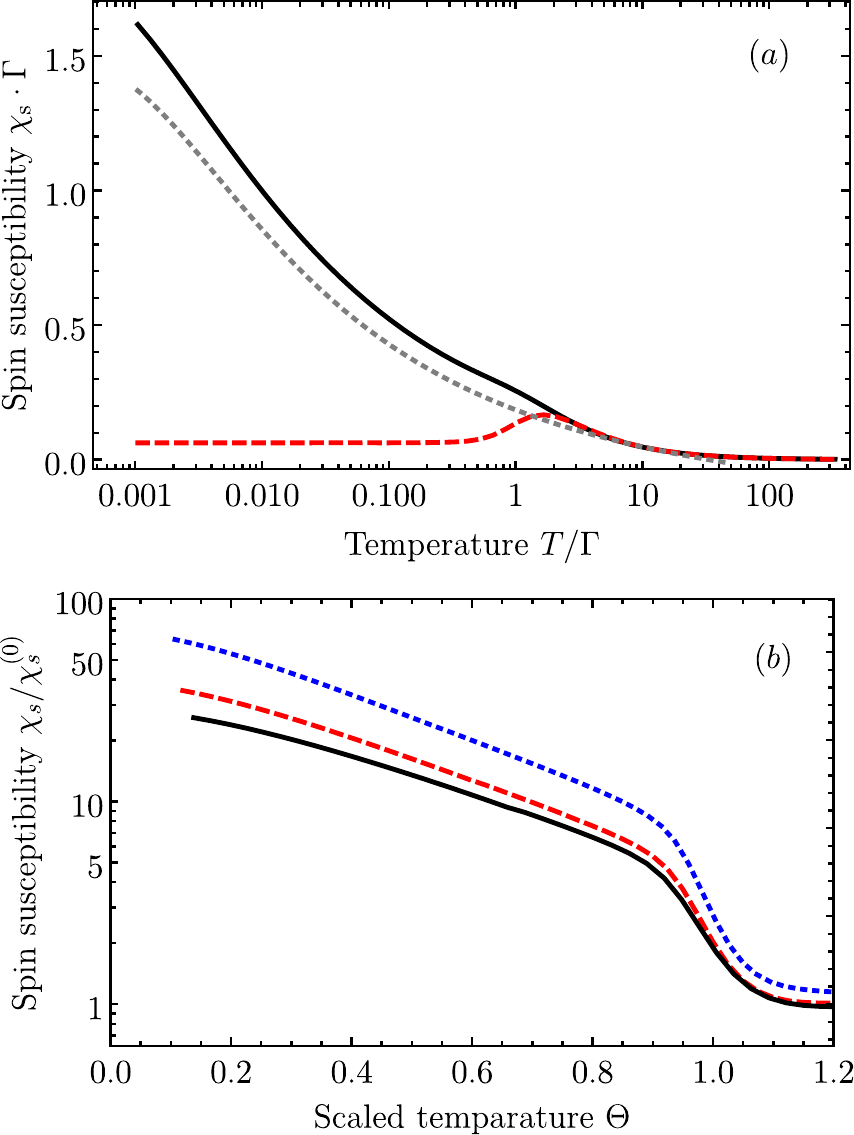}
  \caption{(a) The spin susceptibility as a function of temperature calculated numerically (black solid curve), analytically (gray dotted curve), and in the Hartree--Fock approximation (red dashed curve) for the same parameters as in Fig.~\ref{fig:Polarization_eV_DOS}. (b)~The ratio of the spin susceptibilities calculated with and without many body correlations, $\chi_s/\chi_s^{(0)}$, as a function of the scaled temperature $\Theta$ [Eq.~\eqref{eq:t}] for the same parameters as in Fig.~\ref{fig:Polarization_eV_DOS} except for the larger ratios of the Fermi energy and the band width $(E_F-E_0)/W=0.13$ and $0.4$ for the red dashed and blue dotted curves, respectively.}
  \label{fig:Polarization_T}
\end{figure}

This suggests the introduction of the spin susceptibility $\chi_s$ for the current induced spin accumulation as
\begin{equation}
  \chi_s=\lim_{V\to 0}\frac{P}{eV}.
\end{equation}
Its temperature dependence is shown in Fig.~\ref{fig:Polarization_T}(a) by the black curve for the same parameters as in Fig.~\ref{fig:Polarization_eV_DOS}. One can see that it strongly increases with decrease of temperature. Note that our approach is valid as long as $T>T_K$ only, and $T_K=3.5\cdot10^{-4}\,\Gamma$ for these parameters.

The gray dashed curve represents an analytical approximation, which is quite cumbersome and is given in the Supplemental Material~\cite{supp}. Its analysis for low temperatures $T\gtrsim T_K$ shows that the spin susceptibility can be estimated as
\begin{equation}
  \label{eq:est}
  \chi_s\sim\frac{1}{\Gamma}\times\frac{E_F-E_0}{\Gamma}.
\end{equation}
Here the first factor reflects the fact that the density of states has a maximum at the Fermi energy of the order of $1/\Gamma$ caused by the many body correlations (Kondo peak), so the difference of the occupancies in the leads sharply affects the occupancies of the spin states in the QD. The second factor is related to the interaction induced instability: once the spin up electron enters the QD, it suppresses the tunneling of the spin down electron, so the spin polarization increases.

For comparison, the red dashed curve in Fig.~\ref{fig:Polarization_T}(a) shows the spin susceptibility $\chi_s^{(0)}$ calculated neglecting the many body correlations (Hartree-Fock approximation)~\cite{supp}. At high temperatures $T\gtrsim\Gamma$ it coincides with the black curve. However, at low temperatures, $T\ll\Gamma$, this approximation strongly underestimates the spin susceptibility and gives $\chi_s\sim1/(E_F-E_0)$.

To underline the role of many body correlations in the effect of current induced spin accumulation, we plot the ratio $\chi_s/\chi_s^{(0)}$ in Fig.~\ref{fig:Polarization_T}(b) as a function of scaled temperature
\begin{equation}
  \label{eq:t}
  \Theta=\frac{\ln(T/T_K)}{\ln(\Gamma/T_K)}.
\end{equation}
This parameter equals zero when $T=T_K$, equals unity when $T=\Gamma$ and logarithmically scales in between (note that our approach is valid for $\Theta>0$ only). The black curve in Fig.~\ref{fig:Polarization_T}(b) is calculated for the same parameters as Fig.~\ref{fig:Polarization_T} and shows that the ratio $\chi_s/\chi_s^{(0)}$ exceeds 10 when temperature approaches the Kondo temperature. The red dashed and blue dotted curves are calculated for the chiral quasi bound state located deeper below the Fermi energy, i.e. larger $(E_F-E_0)/W$. For simplicity of the numerical calculations, we tuned this dimensionless parameter by decreasing the band width $W$. One can see that the deeper the localization (or the larger the Fermi energy, or the smaller the band width) the larger role of the many body correlations. In particular, the increase of the spin susceptibility due to them approaches 100 at low temperatures, as shown by the blue dotted curve. This is the main result of this work.

\section{Discussion}Above we considered a completely chiral quasi bound state (i.e. $V_{-|k|,+}=V_{|k|,-}=0$). Generally, for arbitrary bias the results do not change qualitatively for the weaker chirality, but the current induced spin accumulation decreases. In particular case of small bias, the spin susceptibility is simply proportional to the chirality, $\chi_s\propto\mathcal C=\left(|V_{k_0,+}^2|-|V_{k_0,-}^2|\right)/\left(|V_{k_0,+}^2|+|V_{k_0,-}^2|\right)$~\cite{supp}.

As an outlook, we believe that the predicted enhancement of the current induced spin accumulation by the many body correlations is a general phenomenon. So it would be important to apply other theoretical approaches such as the numerical renormalization group method~\cite{PhysRevB.84.193411,PhysRevLett.108.046601,PhysRevB.93.075148}, to establish scaling relations in the strong coupling regime at temperatures below $T_K$, and to study the manifestations of the current induced spin accumulation in the transport properties. Qualitatively, we note that the completely chiral quasi bound state does not lead to the back scattering of the particles in the quantum wire, because each spin state is coupled to the particles propagating only in one direction. Thus the increase of chirality $C$ leads to the disappearance of the Kondo resonance in the differential conductivity. In addition, the current induced spin accumulation in the QD can be probed optically by means of the polarized photoluminescence and spin induced Faraday rotation of the probe light, or electrically using point contacts~\cite{debray2009all,chuang2015all} and scanning tunneling microscopy~\cite{spin_STM}. We note that the Onsager relations also imply that the spin pumping of the QD by external means would lead to the electric current along the quantum wire.

The proposed model allows for various generalizations, which can be studied theoretically and experimentally. For example, it would be interesting to take into account interactions between the particles in the quantum wire or consider a few QDs side coupled to the quantum wire. The spin accumulation in the QD can be also induced by the different temperatures in the leads similar to the spin Nernst effect.

The typical parameters of the system for the experimental realization of the current induced spin accumulation are the lengths of the order of $100$~nm and the coupling strength $\Gamma\sim10~\mu$eV. Then, for example, for the parameters used in Fig.~\ref{fig:Polarization_eV_DOS} we obtain $T_K\sim50~\mu$K, which is quite low. We note however, that the many body correlations significantly enhance the spin polarization even at the temperatures smaller than, but comparable to $\Gamma\sim100$~mK, which is easier to reach.

Indeed, a very similar system was recently realized experimentally~\cite{PhysRevLett.128.027701}. However, the QD was placed inside the quantum wire, so the quasi bound state was not chiral and the effect of the current induced spin accumulation was symmetry forbidden. If QD is shifted along $y$ direction, our theory predicts significant spin polarization increased by the Kondo effect in this system.

\section{Conclusion}We have demonstrated that the Kondo effect can be exploited to enhance the spin galvanic effects such as the current induced spin accumulation. In particular, the many body correlations are shown to parametrically increase the degree of the spin polarization in chiral quasi bound state in the QD side coupled to the quantum wire. The spin susceptibility to the electric current can be enhanced by almost two orders of magnitude for the realistic system parameters.



We thank \href{http://www.ioffe.ru/theory/sector/krainov.html}{I. V. Krainov} for fruitful discussions and the Foundation for the Advancement of Theoretical Physics and Mathematics ``BASIS''. Analytical calculations by D.S.S. were supported by the Russian Science Foundation grant No. 21-72-10035. V.N.M. acknowledges support from the Interdisciplinary Scientific and Educational School of Moscow University ``Photonic and Quantum technologies. Digital medicine''.

\renewcommand{\i}{\ifr}
\let\oldaddcontentsline\addcontentsline
\renewcommand{\addcontentsline}[3]{}


%

\let\addcontentsline\oldaddcontentsline
\makeatletter
\renewcommand\tableofcontents{%
    \@starttoc{toc}%
}
\makeatother
\renewcommand{\i}{{\rm i}}


\onecolumngrid
\vspace{\columnsep}
\begin{center}
\newpage
\makeatletter
{\large\bf{Supplemental Material:\\\@title}}
\makeatother
\end{center}
\vspace{\columnsep}

\twocolumngrid
Supplemental Material includes the following topics:\\

\hypersetup{linktoc=page}
\tableofcontents
\vspace{\columnsep}

\counterwithin{figure}{section}
\renewcommand{\thesection}{S\arabic{section}}
\renewcommand{\section}[1]{\oldsec{#1}}
\renewcommand{\thepage}{S\arabic{page}}
\renewcommand{\theequation}{S\arabic{equation}}
\renewcommand{\thefigure}{S\arabic{figure}}
\renewcommand{\bibnumfmt}[1]{[S#1]}
\renewcommand{\citenumfont}[1]{S#1}

\setcounter{page}{1}
\setcounter{section}{0}
\setcounter{equation}{0}
\setcounter{figure}{0}

\section{S1. Derivation of Green's functions}

Here we present for the completeness the derivation of the retarded Green's functions for the Anderson Hamiltonian. We follow the approach of Ref.~\onlinecite{S_PhysRevLett.70.2601}, which is explained in more detail in Ref.~\onlinecite{S_haug2008quantum}.

We start from the Hamiltonian [Eq.~\eqref{eq:Ham_main} in the main text]
\begin{multline}
  \label{eq:Ham_S}
  \mathcal H=E_0\sum_\pm n_\pm+Un_+n_-+\sum_{k,\pm}E_kn_{k,\pm}\\
  +\sum_{k,\pm}\left(V_{k,\pm}d_\pm^\dag c_{k,\pm}+{\rm H.c.}\right),
\end{multline}
where $E_0$ is the localization energy in the QD, $n_\pm=d_\pm^\dag d_\pm$ are the occupancies of the spin up and spin down states in the quantum dot (QD) with $d_\pm$ ($d_\pm^\dag$) being the corresponding annihilation (creation) operators, $U$ is the Coulomb repulsion energy in the QD, $E_k$ is the spin independent dispersion of the states in the quantum wire with $k$ being the wave vector along the wire, $n_{k,\pm}=c_{k,\pm}^\dag c_{k,\pm}$ are the occupancies of the spin up and spin down states with the wave vector $k$ in the wire with $c_{k,\pm}$ ($c_{k,\pm}^\dag$) being the corresponding annihilation (creation) operators, and finally $V_{k,\pm}$ are the spin dependent tunneling matrix elements between the quantum wire and the QD. The time inversion symmetry implies that
\begin{equation}
  \label{eq:time}
  V_{k,+}=V_{-k,-}^*.
\end{equation}

We note that for the previously investigated case of the hole in the complex valence band~\cite{S_Mantsevich2022}, the subscript $\pm$ for the states in the wire refers to the heavy holes with the spin $\pm3/2$ along the structure growth axis (perpendicular to the plain containing the quantum wire and the QD), and it refers to the light holes with the spin $\mp1/2$ in the QD along the same axis.

The retarded Green's functions can be obtained from the following Heisenberg equations for the operators (${\hbar=1}$):
\begin{subequations}
  \label{eq:sys}
  \begin{equation}
    \label{eq:d}
    \i\frac{\d d_\pm}{\d t}=E_0 d_\pm+U d_\pm n_{\mp}+\sum_kV_{k,\pm}^*c_{k,\pm},
  \end{equation}
  \begin{equation}
    \label{eq:c}
    \i\frac{\d c_{k,\pm}}{\d t}=E_k c_{k,\pm}+V_{k,\pm}d_\pm
  \end{equation}
  \begin{multline}
    \label{eq:cn}
    \i\frac{\d(d_\pm n_{\mp})}{\d t}=(E_0+U)d_\pm n_{\mp}+\sum_k\left(V_{k,\pm}^*c_{k,\pm}n_{\mp}\right.\\\left.+V_{k,\mp}c_{k,\mp}^\dag d_{\pm} d_{\mp}-V_{k,\mp}^*c_{k,\mp}d_{\mp}^\dag d_{\pm}\right),
  \end{multline}
  \begin{multline}
    \label{eq:c1}
    \i\frac{\d(c_{k,\pm}n_{\mp})}{\d t}=E_kc_{k,\pm}n_{\mp}+V_{k,\pm}d_\pm n_{\mp}\\+\sum_q\left(-V_{q,\mp}c_{k,\pm}c_{q,\mp}^\dag d_{\mp}+V_{q,\mp}^*c_{k,\pm}d_{\mp}^\dag c_{q,\mp} \right),
  \end{multline}
  \begin{multline}
    \label{eq:c2}
    \i\frac{\d(c_{k,\mp}^\dag d_{\pm} d_{\mp})}{\d t}=(2E_0+U-E_k)c_{k,\mp}^\dag d_{\pm} d_{\mp}+V_{k,\mp}^*d_\pm n_{\mp}\\+\sum_q\left(V_{q,\pm}^*c_{k,\mp}^\dag c_{q,\pm} d_{\mp}+V_{q,\mp}^*c_{k,\mp}^\dag d_{\pm} c_{q,\mp}\right),
  \end{multline}
  \begin{multline}
    \label{eq:c3}
    \i\frac{\d(c_{k,\mp}d_{\mp}^\dag d_{\pm})}{\d t}=E_kc_{k,\mp}d_{\mp}^\dag d_{\pm}+V_{k,\mp}\left(d_\pm-d_\pm n_{\mp}\right)\\+\sum_q\left(-V_{q,\mp}c_{k,\mp}c_{q,\mp}^\dag d_{\pm}+V_{q,\pm}^*c_{k,\mp}d_{\mp}^\dag c_{q,\pm}\right).
  \end{multline}
\end{subequations}
These equations allow one to calculate in the steady state the correlation functions of the operators $A$ and $B$ like
\begin{equation}
  \braket{A,B}^R(t)\equiv-\i\braket{A(t)B+BA(t)}\theta(t)
\end{equation}
with $\theta(t)$ being the Heaviside step function from the equations
\begin{equation}
  \label{eq:AB}
  \frac{\d}{\d t}\braket{A,B}^R(t)=-\i\braket{AB+BA}\delta(t)+\braket{\frac{\d A}{\d t},B}^R(t),
\end{equation}
where $\delta(t)$ is the Dirac delta function.

Our aim is to calculate the retarded spin dependent Green's functions
\begin{equation}
  G_\pm^R(t)\equiv\braket{d_\pm,d_\pm^\dag}^R.
\end{equation}
Thus we consider $B=d_\pm^\dag$ in Eq.~\eqref{eq:AB} and different operators $A$ from Eqs.~\eqref{eq:sys}.

In the Hartree--Fock approximation one uses Eqs.~(\ref{eq:d}---\ref{eq:cn}), truncates the system using the approximations
\begin{equation}
  \label{eq:HF_approx}
  c_{k,\pm}n_\mp=c_{k,\pm}\braket{n_\mp},
  \qquad
  c_{k,\mp}^\dag d_{\pm} d_{\mp}=c_{k,\mp}d_{\mp}^\dag d_{\pm}=0,
\end{equation}
and uses the following correlators
\begin{multline}
  \braket{d_\pm^\dag d_\pm+d_\pm d_\pm^\dag}=1,
  \qquad
  \braket{d_\pm^\dag c_{k,\pm}+c_{k,\pm}d_\pm^\dag}=0,
  \\
  \braket{d_\pm^\dag d_\pm n_\mp+d_\pm n_\mp d_\pm^\dag}=\braket{n_\mp},
\end{multline}
where $\braket{n_\pm}$ are the steady state occupancies of the spins states of the QD, which should be calculated self consistently. In this way, we obtain the Green's functions in the limit of strong Coulomb interaction ($U\to\infty$)~\cite{S_haug2008quantum}
\begin{equation}
  \label{eq:GR_HF}
  G_\pm^{R(0)}=\frac{1-\braket{n_\mp}}{\omega-E_0-\Sigma_{0}^R(1-\braket{n_\mp})},
\end{equation}
where the self energy
\begin{equation}
  \label{eq:Sigma0R}
  \Sigma_0^R=\sum_{k}\frac{|V_{k,\pm}^2|}{\omega-E_k+\i 0}
\end{equation}
is spin independent due to Eq.~\eqref{eq:time}.

The Hartree--Fock approximation does not capture the Kondo effect, therefore, it is necessary to go beyond Eq.~\eqref{eq:HF_approx} and consider additional Eqs.~(\ref{eq:c1}---\ref{eq:c3}). To truncate the system one uses the approximations
\begin{multline}
  \label{eq:Kondo_truncation}
  c_{k,\mp}^\dag d_{\pm} c_{q,\mp}=-\delta_{k,q}f_kd_\pm,
  \quad
  c_{k,\mp}c_{q,\mp}^\dag d_{\pm}=\delta_{k,q}(1-f_k)d_\pm,
  \\
  c_{k,\pm}c_{q,\mp}^\dag d_{\mp}=c_{k,\pm}d_{\mp}^\dag c_{q,\mp}=c_{k,\mp}^\dag c_{q,\pm} d_{\mp}=c_{k,\mp}d_{\mp}^\dag c_{q,\pm}=0,
\end{multline}
where ($k_B=1$)
\begin{subequations}
\begin{equation}
  f_{|k|}=\frac{1}{1+\exp\left[(E_k-E_F^L)/T\right]}
\end{equation}
and
\begin{equation}
  f_{-|k|}=\frac{1}{1+\exp\left[(E_k-E_F^R)/T\right]}
\end{equation}
\end{subequations}
are the spin independent occupancies of the states with the wave vector $|k|$ ($-|k|$) in the quantum wire with $E_F^L$ ($E_F^R$) being the Fermi level in the left (right) lead. One also uses the following additional approximations for the correlation functions:
\begin{multline}
  \label{eq:Kondo_corr}
  \braket{d_\pm^\dag c_{k,\mp}^\dag d_{\pm} d_{\mp}+c_{k,\mp}^\dag d_{\pm} d_{\mp}d_\pm^\dag}\\=\braket{d_\pm^\dag c_{k,\mp}d_{\mp}^\dag d_{\pm}+c_{k,\mp}d_{\mp}^\dag d_{\pm}d_\pm^\dag}=0.
\end{multline}
In this way, we obtain the Green's functions in the limit of large $U$~\cite{S_PhysRevLett.70.2601}:
\begin{equation}
  \label{eq:GR_S}
  G_\pm^{R}=\frac{1-\braket{n_\mp}}{\omega-E_0-\Sigma_{0}^R-\Sigma_{1,\pm}^R},
\end{equation}
where
\begin{equation}
  \label{eq:S1}
  \Sigma_{1,\pm}^R=\sum_{k}\frac{f_k|V_{k,\mp}^2|}{\omega-E_k+\i 0}.
\end{equation}
These expressions take into account the many body correlations in the minimal approximation, which allows one to account for the Kondo effect.

\section{S2. Wide band approximation}

To obtain transparent expressions for the Green's functions we consider the wide band approximation, which is given by the substitution
\begin{multline}
  \sum_{k}|V_{k,\pm}^2|\phi(k)\to\frac{\Gamma}{\pi}\int\limits_{-W}^W\left[\frac{1\pm\mathcal C}{2}\phi(|k|)\right.\\\left.+\frac{1\mp\mathcal C}{2}\phi(-|k|)\right]\d E_k,
\end{multline}
where $\phi(k)$ stands for arbitrary function, $W$ is the band width, $\Gamma=\pi D(E_0)(|V_{k_0,\pm}^2+|V_{-k_0,\pm}^2|)/4$ is the spin independent width of the quasi bound state with $D(E_0)$ being the density of states in the quantum wire including spin and the two directions of the propagation and $k_0$ being the wave vector corresponding to the energy of the quasi bound state so that $E_{k_0}=E_0$ ($k_0>0$), and
\begin{equation}
  \mathcal C=\frac{|V_{k_0,+}^2|-|V_{k_0,-}^2|}{|V_{k_0,+}^2|+|V_{k_0,-}^2|}
\end{equation}
is the chirality of the quasi bound state. The band width $W$ is assumed to be the largest energy scale (except for $U$).

With this substitution one obtains from Eq.~\eqref{eq:Sigma0R} the self energy
\begin{equation}
  \label{eq:S0_wide}
  \Sigma_0^R=-\i\Gamma.
\end{equation}
Then the Green's function in the Hartree--Fock approximation, Eq.~\eqref{eq:GR_HF}, reads
\begin{equation}
  \label{eq:GR_HF_wide}
  G_\pm^{R(0)}=\frac{1-\braket{n_\mp}}{\omega-E_0+\i\Gamma(1-\braket{n_\mp})}.
\end{equation}

Further, another self energy, Eq.~\eqref{eq:S1}, has the form
\begin{multline}
  \label{eq:S1_wide}
  \Sigma_{1,\pm}^R=\frac{\Gamma}{2\pi}\int\limits_{-W}^W\frac{(1\mp\mathcal C)f_L(E)+(1\pm\mathcal C)f_R(E)}{\omega-E+\i 0}\d E\\\equiv\Sigma_{1,\pm}^{R,L}+\Sigma_{1,\pm}^{R,R},
\end{multline}
where
\begin{equation}
  f_{L/R}(E)=\frac{1}{1+\exp\left[(E-E_F^{L/R})/T\right]}
\end{equation}
[Eq.~\eqref{eq:fLR} in the main text] and we have separated the two contributions $\Sigma_{1,\pm}^{R,L/R}$ related to the left/right leads. One can see that these self energies logarithmically diverge for large $W$ because of the contribution from the low frequencies. This divergence determines the Kondo temperature
\begin{equation}
  T_K=W\exp(-\pi|E_F-E_0|/\Gamma)
\end{equation}
in this model with $E_F=(E_F^L+E_F^R)/2$, which should be smaller than $T$ to ensure the validity of the approach: $T_K<T$. The Green's functions, Eq.~\eqref{eq:GR_S}, then read
\begin{equation}
  \label{eq:GR_wide}
  G_\pm^{R}=\frac{1-\braket{n_\mp}}{\omega-E_0+\i\Gamma-\Sigma_{1,\pm}^R}.
\end{equation}

\begin{figure}
  \includegraphics[width=0.95\linewidth]{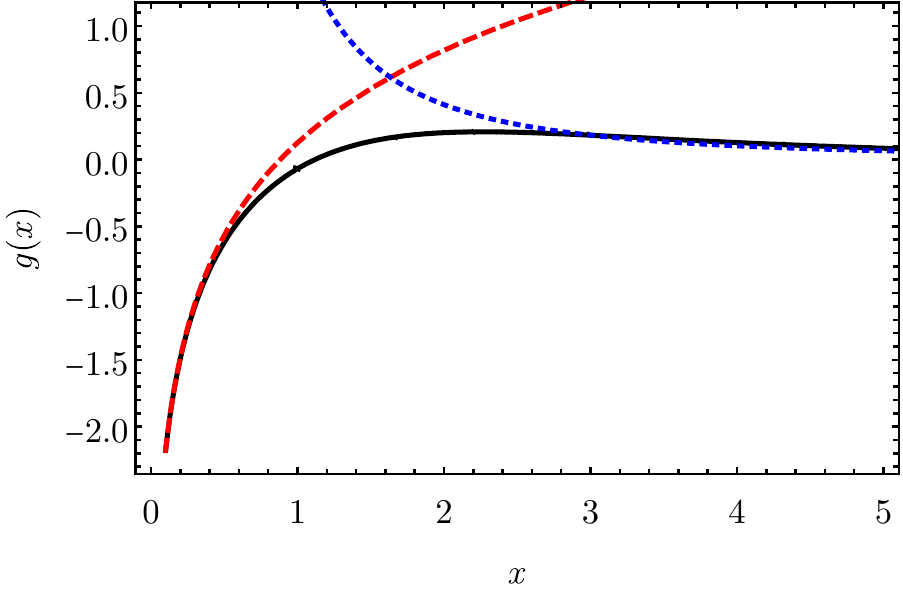}
  \caption{Function $g(x)$ given by Eq.~\eqref{eq:g} (black solid curve) and its asymptotes Eq.~\eqref{eq:g_small} (red dashed curve) and Eq.~\eqref{eq:g_large} (blue dotted curve).}
  \label{fig:g_x}
\end{figure}

For relatively low temperatures, $T\ll\Gamma$ the integrals in Eq.~\eqref{eq:S1_wide} yield
\begin{multline}
  \label{eq:S1_anal}
  \Sigma_{1,\pm}^{L/R}=\frac{\Gamma_\mp^{L/R}}{\pi}\left[\ln\left|\frac{W}{\omega-E_F^{L/R}}\right|+g\left(\frac{|\omega-E_F^{L/R}|}{T}\right)\right.\\\left.-\i\pi f_{L/R}(\omega)\right],
\end{multline}
where
\begin{equation}
  \Gamma_\pm^L=\frac{1\pm\mathcal C}{2}\Gamma,
  \qquad
  \Gamma_\pm^R=\frac{1\mp\mathcal C}{2}\Gamma,
\end{equation}
and
\begin{equation}
  \label{eq:g}
  g(x)=\int\limits_0^\infty\frac{\d y}{y}\left[\frac{\sh(y)}{\ch(y)+\ch(x)}-\theta(y-x)\right].
\end{equation}
This dimensionless function is shown in Fig.~\ref{fig:g_x} by black solid line. For small $x$ it has an asymptote
\begin{equation}
  \label{eq:g_small}
  g(x)=\ln(2x/\pi)+\gamma
\end{equation}
with $\gamma\approx0.58$ being the Euler constant. For large $x$ the asymptote reads
\begin{equation}
  \label{eq:g_large}
  g(x)=\pi^2/(6x^2).
\end{equation}
These asymptotes are shown in Fig.~\ref{fig:g_x} by red dashed and blue dotted curves and cover almost the whole range of $x$ from zero to infinity.

At relatively high temperatures $T\sim\Gamma$ one has to use Eq.~\eqref{eq:S1_wide} to calculate the Green's functions.

\section{S3. Calculation of spin susceptibility}

To calculate the current induced spin accumulation in the QD, we write the electric current of the given spin component to the QD from the left/right leads~\cite{S_haug2008quantum}:
\begin{equation}
  J_\pm^{L/R}=e\int\frac{\d\omega}{2\pi}\left[\i\Gamma_\pm^{L/R}G_\pm^<-2f_{L/R}\Gamma_\pm^{L/R}\Im G_\pm^R \right],
\end{equation}
where $G_\pm^<$ are the lesser Green's functions of the QD. In the steady state, due to the spin and charge conservation in the QD, we have
\begin{equation}
  \label{eq:J_tot}
  J_\pm^L+J_\pm^R=0.
\end{equation}
From this relation we obtain the occupancies of the spin states of the QD in the form
\begin{multline}
  \label{eq:npm}
  \braket{n_\pm}=-\i\int\frac{\d\omega}{2\pi}G_\pm^<\\=-\int\frac{\d\omega}{\pi\Gamma}\left[f_L\Gamma_\pm^L+f_R\Gamma_\pm^R\right]\Im G_\pm^R.
\end{multline}
These occupancies define the spin polarization in the QD even beyond the linear response regime.

We note that the lesser Green's functions can be also calculated using the analytical continuation method~\cite{S_Langreth1976,S_kadanoff2018quantum,S_haug2008quantum} following the approach of Refs.~\onlinecite{S_Niu_1999,S_PhysRevB.68.195318}. However, the phenomenological truncation of the Heisenberg equations using the relations~\eqref{eq:Kondo_truncation} and~\eqref{eq:Kondo_corr} yields in this way the Green's functions $G_\pm^<$ that do not coincide with $-2\i f_{L/R}G_\pm^R$ in the equilibrium. Therefore, for consistency we prefer to determine the occupancies of the QD states from the charge and spin conservation relations~\eqref{eq:J_tot}.

From Eq.~\eqref{eq:GR_wide} one can see that Eq.~\eqref{eq:npm} can be rewritten as
\begin{equation}
  \label{eq:npm_Ipm}
  \braket{n_\pm}=I_\pm(1-\braket{n_\mp}),
\end{equation}
where
\begin{equation}
  \label{eq:Ipm}
  I_\pm=-\int\frac{\d\omega}{\pi\Gamma}\left[f_L\Gamma_\pm^L+f_R\Gamma_\pm^R\right]\Im\frac{1}{\omega-E_0-\Sigma_{0}^R-\Sigma_{1,\pm}^R}
\end{equation}
do not depend on $\braket{n_\pm}$ (here we take into account that for finite $W$ Eq.~\eqref{eq:S0_wide} may be weakly violated). From the solution of these equations we obtain the spin polarization in the QD
\begin{equation}
  P\equiv\frac{\braket{n_+-n_-}}{\braket{n_++n_-}}=\frac{I_+-I_-}{I_++I_--2I_+I_-}
\end{equation}
and its occupancy
\begin{equation}
  \label{eq:n_tot}
  \braket{n_++n_-}=\frac{I_++I_--2I_+I_-}{1-I_+I_-}.
\end{equation}
These expressions generally define the spin state of the QD.

To calculate the spin susceptibility, we consider the symmetrically applied bias $eV$: $E_F^L=E_F+eV/2$, $E_F^R=E_F-eV/2$ and set $E_F=0$ to be specific. In the case of $eV=0$ one has $I_+=I_-=I_0$, where
\begin{equation}
  \label{eq:I00}
  I_0=-\int\frac{\d\omega}{\pi}\Im\frac{f_0(\omega)}{\omega-E_0-\Sigma_0^R-\Sigma_{1,0}^R}
\end{equation}
with
\begin{equation}
  f_0(E)=\frac{1}{1+\exp\left[(E-E_F)/T\right]}
\end{equation}
and
\begin{equation}
  \Sigma_{1,0}^R=\Gamma\int\limits_{-W}^W\frac{\d E}{\pi}\frac{f_0(E)}{\omega-E+\i 0}.
\end{equation}
In the first order in $eV$ we obtain
\begin{multline}
  \label{eq:Delta_I}
  I_+-I_-=-\mathcal C\int\frac{\d\omega}{\pi}\Im\left[\frac{f_L(\omega)-f_R(\omega)}{\omega-E_0-\Sigma_0^R-\Sigma_{1,0}^R}\right.\\\left.
    -\frac{\Gamma f_0(\omega)}{(\omega-E_0-\Sigma_0^R-\Sigma_{1,0}^R)^2}\int\limits_{-W}^W\frac{\d E}{\pi}\frac{f_L(E)-f_R(E)}{\omega-E+\i 0}\right].
\end{multline}
Then the spin susceptibility is given by
\begin{equation}
  \label{eq:chi_I}
  \chi_s=\frac{(I_+-I_-)/(eV)}{2I_0(1-I_0)}.
\end{equation}
From Eq.~\eqref{eq:Delta_I} one can see that it linearly depends on the chirality $\mathcal C$, so we focus on the case of $\mathcal C=1$ in what follows and in the main text.

\section{S4. Estimation of spin susceptibility}

To obtain a qualitative understanding of the current induced spin accumulation enhancement due to Kondo effect, we derive an analytical approximation for the spin susceptibility at low temperatures, $T_K<T\ll\Gamma$ and at high Fermi level $E_F-E_0\gg\Gamma$.

First of all, we note that in this limit $\braket{n_++n_-}\approx 1$ because of the strong Coulomb interaction, so from Eq.~\eqref{eq:n_tot} we obtain $I_0\approx 1$. Then we note that the denominator in Eq.~\eqref{eq:chi_I} is close to zero, so the spin susceptibility $\chi_s$ is large. This is related to the fact that for the large Coulomb interaction the system is a sort of unstable: a small occupancy of one spin state strongly suppresses occupancy of another at the same moment.

To estimate $1-I_0$ we approximate Eq.~\eqref{eq:I00} as follows:
\begin{multline}
  \label{eq:I0}
  I_0\approx-\int\limits_{-W}^{E_F}\frac{\d\omega}{\pi}\Im\frac{1}{\omega-E_0+\i\Gamma}\\=\frac{1}{\pi}\left[\arctg\left(\frac{W+E_0}{\Gamma}\right)+\arctg\left(\frac{E_F-E_0}{\Gamma}\right)\right].
\end{multline}

Then we rewrite Eq.~\eqref{eq:Ipm} for the case of $\mathcal C=1$ as
\begin{equation}
  \label{eq:Ipm1}
  I_{+/-}=-\int\frac{\d\omega}{\pi}\Im\frac{f_{L/R}}{\omega-E_0-\Sigma_{0}^R-\Sigma_{1,+/-}^R}.
\end{equation}
Further from Eqs.~\eqref{eq:Sigma0R} and~\eqref{eq:S1_wide} we obtain
\begin{equation}
  \Im\Sigma_0^R=-\Gamma,
  \qquad
  \Im\Sigma_{1,+/-}^R=-\Gamma f_{R/L},
\end{equation}
thus we arrive at
\begin{equation}
  I_{+/-}=\int\frac{\d\omega}{\pi}\frac{\Gamma f_{L/R}(1+f_{R/L})}{\left|\omega-E_0-\Sigma_{0}^R-\Sigma_{1,+/-}^R\right|^2}.
\end{equation}

Next we separate the three contributions to the current induced spin accumulation as follows [cf. Eq.~\eqref{eq:Delta_I}]:
\begin{equation}
  \label{eq:3contrib}
  I_+-I_-=\Delta I_1+\Delta I_2+\Delta I_3,
\end{equation}
where
\begin{subequations}
  \begin{equation}
    \Delta I_1=\int\frac{\d\omega}{\pi}\frac{\Gamma(f_{L}-f_R)}{\left|\omega-E_0-\Sigma_{0}^R-\Sigma_{1,0}^R\right|^2},
  \end{equation}
  \begin{equation}
    \Delta I_2=\int\frac{\d\omega}{\pi}\frac{2\Gamma^3(f_{L}-f_R)f_0(1+f_0)^2}{\left|\omega-E_0-\Sigma_{0}^R-\Sigma_{1,0}^R\right|^4},
  \end{equation}
  \begin{multline}
    \Delta I_3=\int\frac{\d\omega}{\pi}2\Gamma^2f_0(1+f_0)\Re(\Sigma_{1,+}^R-\Sigma_{1,-}^R)\\\times\frac{\Re(\omega-E_0-\Sigma_{0}^R-\Sigma_{1,0}^R)}{\left|\omega-E_0-\Sigma_{0}^R-\Sigma_{1,0}^R\right|^4}.
  \end{multline}
\end{subequations}
In the first two contributions, we replace $f_L-f_R$ with $eV\delta(\omega-E_F)$ and note that at $\omega=E_F$ it follows from Eqs.~\eqref{eq:S1_anal} and~\eqref{eq:g_small} that
\begin{equation}
  \Sigma_{1,0}^R=\frac{\Gamma}{\pi}\left[\ln\left(\frac{2W}{\pi T}\right)+\gamma\right]-\i\frac{\Gamma}{2}
\end{equation}
and taking into account Eq.~\eqref{eq:S0_wide} we obtain
\begin{equation}
  \label{eq:I1}
  \Delta I_1=\frac{eV}{\pi}\frac{\Gamma}{a^2+(3\Gamma/2)^2},
\end{equation}
\begin{equation}
  \label{eq:I2}
  \Delta I_2=\frac{eV}{\pi}\frac{9\Gamma^3}{4\left[a^2+(3\Gamma/2)^2\right]^2},
\end{equation}
where
\begin{equation}
  a=\left|E_F-E_0-\frac{\Gamma}{\pi}\left[\ln\left(\frac{2W}{\pi T}\right)+\gamma\right]\right|.
\end{equation}
Finally, we estimate the last contribution as
\begin{equation}
  \label{eq:I3}
  \Delta I_3=h(W/T)-h(W/\Gamma),
\end{equation}
where
\begin{equation}
  h(x)=\frac{2/\pi}{\frac{(E_F-E_0)^2}{\Gamma^2}+4-\frac{2(E_F-E_0)}{\pi\Gamma}\ln(x)+\ln^2(x)}.
\end{equation}
Now the analytical estimation for the spin susceptibility is given by Eqs.~\eqref{eq:chi_I}, \eqref{eq:I0}, \eqref{eq:3contrib}, \eqref{eq:I1}, \eqref{eq:I2}, and \eqref{eq:I3}. It is shown by the gray dotted line in Fig.~\ref{fig:Polarization_T}(a) in the main text.

To find an order of magnitude estimation we note that
\begin{equation}
  1-I_0\sim\frac{\Gamma}{E_F-E_0},
  \quad
  \Delta I_{1,2,3}\sim\frac{eV}{\Gamma}.
\end{equation}
So Eq.~\eqref{eq:chi_I} yields [Eq.~\eqref{eq:est} in the main text]
\begin{equation}
  \label{eq:est_S}
  \chi_s\sim\frac{E_F-E_0}{\Gamma^2}.
\end{equation}


For comparison, in the Hartree-Fock approximation we obtain in the same way from Eq.~\eqref{eq:GR_HF_wide} that the occupancies of the spin states obey the equations~\cite{S_Mantsevich2022}
\begin{equation}
  \label{eq:n_HF}
  \braket{n_{+/-}}=\frac{1}{\pi}\int\limits_{-W}^W\frac{\Gamma(1-\braket{n_{-/+}})^2f_{L/R}}{(\omega-E_0)^2+\Gamma^2(1-\braket{n_{-/+}})^2}\d\omega.
\end{equation}
These equations should be solved self consistently for $\braket{n_\pm}$. They were used to calculate the spin susceptibility $\chi_s^{(0)}$ for Fig.~\ref{fig:Polarization_T} in the main text.

In the wide band limit ($W\to\infty$) and for the large Fermi energy ($E_F-E_0\gg\Gamma,T$), Eq.~\eqref{eq:n_HF} can be written in the form of Eq.~\eqref{eq:npm_Ipm} with
\begin{equation}
  I_\pm=1-\frac{\Gamma/\pi}{E_F-E_0}\pm\frac{\Gamma eV}{2\pi(E_F-E_0)^2}.
\end{equation}
Thus, using Eq.~\eqref{eq:chi_I} we find an estimation for the spin susceptibility in the Hartree-Fock approximation:
\begin{equation}
  \chi_s^{(0)}\sim\frac{1}{E_F-E_0}.
\end{equation}
As discussed in the main text, it is parametrically smaller than the spin susceptibility at low temperatures with account for the many body correlations [Eq.~\eqref{eq:est_S}]. The reason for this is the absence of the peak in the density of states at the Fermi energy and underestimation of the role of the Coulomb interaction.

\end{document}